# Crystallization of diamond-like carbon to graphene under low energy ion beam modification

**S.S. Tinchev***
**Institute of Electronics, Bulgarian Academy of Sciences,**
**Sofia 1784, Bulgaria**

**Abstract:**
Low-energy ion beam modification was proposed to create graphene on the top of the insulated diamond-like carbon films. In such low-temperature fabrication process the surface of the amorphous carbon could crystallize to graphene as a result of point defect creation and enhanced diffusion caused by the ion bombardment. In the experiment 130 eV argon ion irradiation was used. After the modification the resistivity of the sample surface drops. Raman spectra of the samples measured at 633 nm showed partial crystallization and were similar to the spectra of defected graphene. This result is very encouraging and we hope that by improving this technology it will be possible to fabricate defect-free graphene, which can be used in electronics without transfer to other substrates

**Introduction**:

Graphene have been the subject of research in the recent years because of its unique electrical, optical and mechanical properties. These properties make them an ideal material for many applications in electronics. Various methods were demonstrated for deposition of large-area graphene films. In the most successfully chemical vapor deposition (CVD) methods such films are deposited on a metal catalytic layer (Ni, Co) or on Cu (using low solubility of carbon in Cu) [1, 2]. However, in the electronics usually graphene sheets on insulating substrates are needed. Therefore the graphene thin films deposited on metals are transferred to insulating substrates, which is a difficult step. Additionally the properties of transferred graphene are deteriorated from large numbers of traps and phonon scattering in typical substrates used like $SiO_2$.

During the writing this paper a publication arise [3], where graphen was transferred to diamond-like carbon substrates. Graphene transistors fabricated from graphen on diamond-like carbon have better properties because of a less impurity and phonon scattering. The disadvantages of graphene transfer, however, still remained.

Here, we present an alternative low temperature method of fabrication of graphene layers on the top of insulated diamond-like carbon films by low energy ion beam irradiation. Generally, ion bombardment causes structural damage in a crystalline material. However, in the ion beam irradiation of amorphous material, ion beam induced epitaxial crystallization of amorphous layers is possible. The mechanism for such crystallization process involves point defect creation and enhanced diffusion caused by ion bombardment.

One should mention that there is a technology for fabrication of graphen without its transfer. In this technology graphene is grown on the semi-insulation surface of SiC by high temperature thermal decomposition of silicon carbide [4]. However, this process requires temperatures over 1000ºC and is not compatible with the current semiconductor technology.

**Calculation of the necessarily ion beam parameters:**

Different ions can be used to modify diamond-like carbon films. Three types of ions were analyzed as possible candidates for low energy ion beam modification of amorphous carbon films. The first two are carbon and hydrogen ions, which are inherent to our a-C:H films and we do not expect to introduce chemical effects. The last one is argon ions, which are widely used in the microelectronic technology and as a noble gas should not react with the carbon.

* e-mail: stinchev@ie.bas.bg

The Monte Carlo SRIM program [5] was used to estimate the necessary energy and doses of the ions in order to modify only about 1 nm on the surface of the amorphous carbon films. Fig. 1 shows the calculated profile of the vacancies produced in amorphous carbon films by 1 keV argon ion irradiation. As expected only the surface of the film is modified. One can see in the Fig.1 also the profile of the implanted argon ions – the curve "Ion range". Obviously the argon ions are implanted far behind the modified surface region and they should not introduce additional effects.

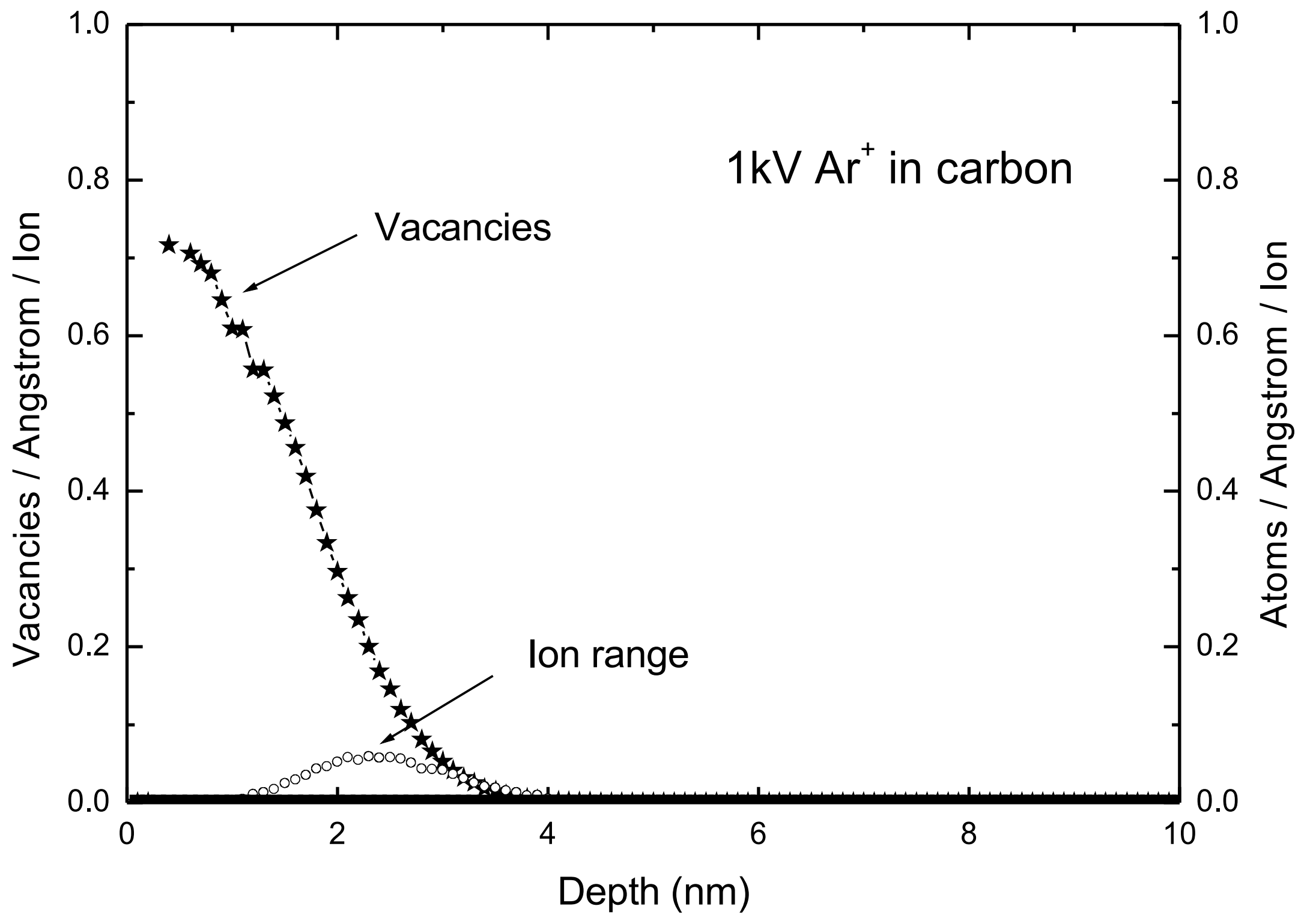

Fig. 1. Calculated vacancies profile and argon ion distribution for 1 keV $Ar^+$ modification of amorphous carbon.

From the maximal number of vacancies ~ 0.7 Vacancies/Angstrom/ion one can estimate the dose needed to break all sp3 bonds in the amorphous carbon films. In the used films they are about 80%. Taking into account that every carbon atom has four bonds we found that the dose should be ~ $4.5 \times 10^{15}$ $Ar^+/cm^2$. This estimated value is in good agreement with the experimental value found in the literature [6].

**Experimental:**

The films used in our experiments were a-C:H films onto single-crystal (100) Si wafers fabricated by PE CVD from benzene vapor diluted with argon. These films can be produced with very different resistivity from soft graphitic-like low resistance films to hard, high resistance films by varying only the bias voltage in a DC PECVD system [7]. As we needed high resistance samples the films used in this work were fabricated at 1kV bias voltage.

The films were modified in DC magnetron system at unipolar pulsed discharges. Pulse biasing of the magnetron is needed because the diamond-like carbon films are highly insulating and ion bombardment with DC voltage would cause charging of the film. The voltage amplitude was 130 V, pulse frequency of 66 kHz and pulse time of 10 μs. The system was evacuated by a combination of diffusion and a mechanical pump. During the modification the pressure of the chamber was 6.6 Pa. The films to be modified were placed in the erosion zone of a 50 mm circular magnetron and the magnetron was specially not water cooled in order to have enhanced temperature and thus enhanced diffusion in the sample during the ion bombardment. The necessary processing time for achieving a necessary dose of ~ $4.5x10^{15}$ $Ar^+/cm^2$ was calculated from the magnetron current and the area of erosion zone assuming homogenous current density distribution in the erosion zone. It was found that the modification time should be shorter then 1 sec.

Electrical resistance and Raman spectra of the films were measured before and after film modification at room temperature. Electrical resistance measurements were performed with 2-point DC and standard lock-in techniques with excitation currents of 10 nA at 469 Hz. Silver paint contacts were applied at the corners of a rectangular sample. Raman spectra were obtained using lasers with 633 nm wavelength.

**Results:**

The virgin diamond-like carbon film was highly insulating. After the ion beam modification its sheet resistivity drops. The measured resistivity of the modified film was ~ 245 kΩ /   . Using ~ 0.34 nm as the estimated thickness of the modified surface layer one can obtain an estimation of the resistivity of the modified material of ~ $8x10^{-5}$ Ω m, which is close to the known resistivity of the graphite [8].

Here it should be noted that resistivity measurement are carried out in air at ambient conditions. It is well known that being in essence a surface graphene is extremely sensitive to contamination. That is actually the reason why graphene can be used as very sensitive gas sensor [9]. In resistivity measurements this phenomenon requires that the graphene samples should be cleaned by annealing in ultrahigh vacuum or in $Ar/H_2$ environment to remove contamination adsorbed on the surface. Another method [10] is current cleaning, which consists of applying a current of several milliamperes through the graphene device at low temperatures. Without such cleaning the adsorbed contamination greatly enhances the measured resistivity of the sample and as a result "real" resistivity of graphene cannot be obtained.

The adsorbed contamination on a grapheme is also the reason for observation of photoelectrical response in single-layer graphene [11]. This phenomenon, initially observed by accident during electrical transport measurements of single-walled carbon nanotubes, is found to be generic for various types of chemisorbed and physisorbed molecules on nanotubes [12]. Actually in this photoelectrical effect a light illumination stimulates molecular desorption (mainly oxygen desorbtion) from the surface and thus the electrical resistance of the sample is changed. Photoelectrical phenomena are also observed with our samples. Our first explanation of this observation was a photoinduced desorption of the oxygen from the sample surface. Later, however, it was found that this is a photoconductivity effect observed also in the pristine DLC film.

Raman measurements were also carried out, because it is well known that Raman spectroscopy is capable to identify graphene and is widely used as an unambiguous method to determine the number of layers of graphene. In our samples, however, there are problems to use Raman spectroscopy. The modified layer is formed on the top of the amorphous carbon film. The penetration depth of the used Raman laser at 633 nm is some hundred nm and the Raman signal is generated practically from the whole film thickness. Therefore a strong Raman signal from underlaying amorphous carbon layer covers the expected G- peak of the graphene - Fig. 2. Nevertheless spectra taken on modified surfaces can be corrected for the underlying amorphous carbon contribution as shown in Fig. 3. Here the spectrum of the virgin film was subtracted from the spectrum of the modified film. In the difference spectrum clear D-peak at 1316 $cm^{-1}$ and G-peak at 1595 $cm^{-1}$ of the modified surface can be recognized. The presence of the D peak indicates the presence of disorder in the modified material.

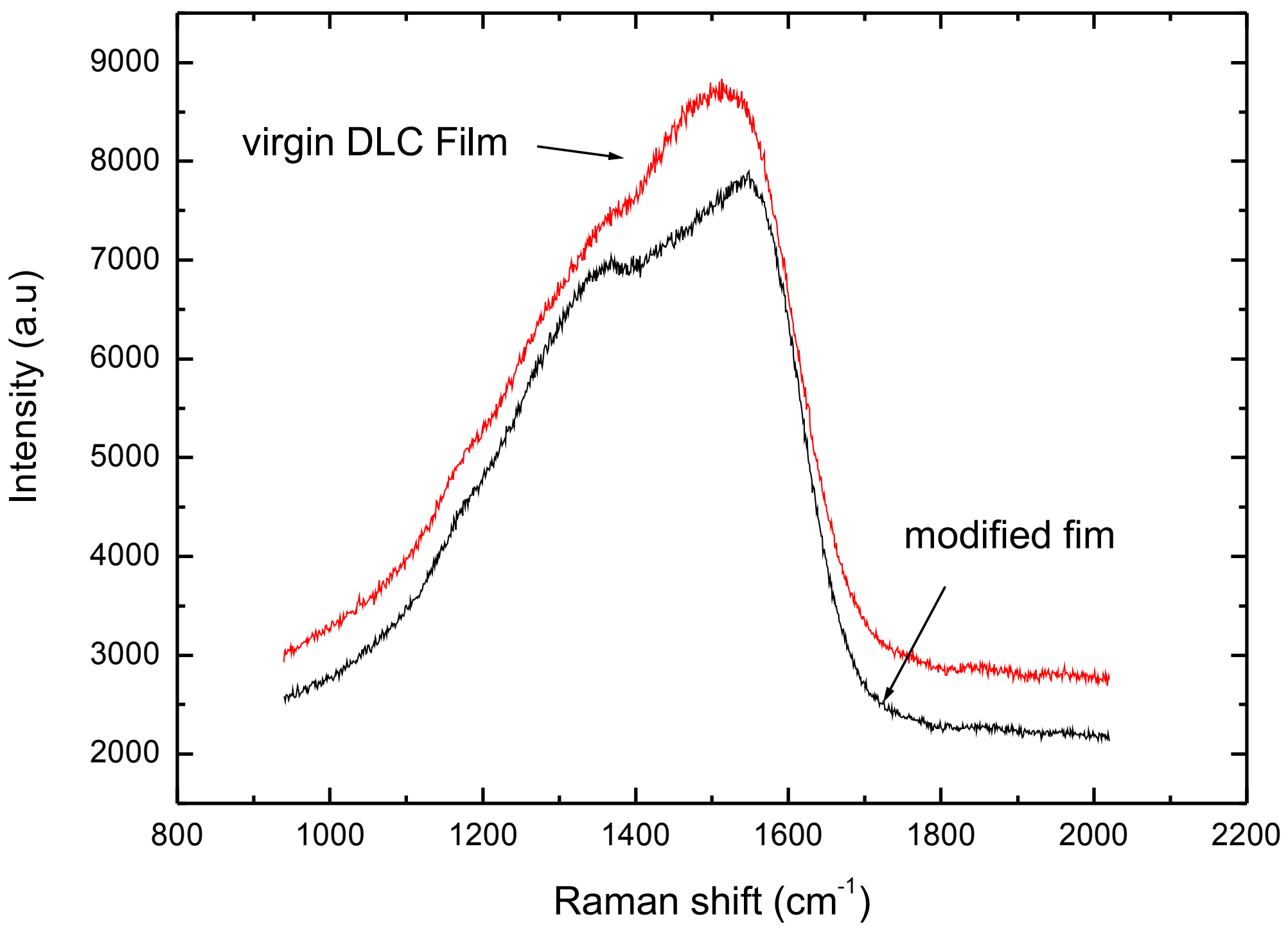

Fig.2 Raman spectra of the diamond-like carbon films before and after the $Ar^{+}$ ion modification.

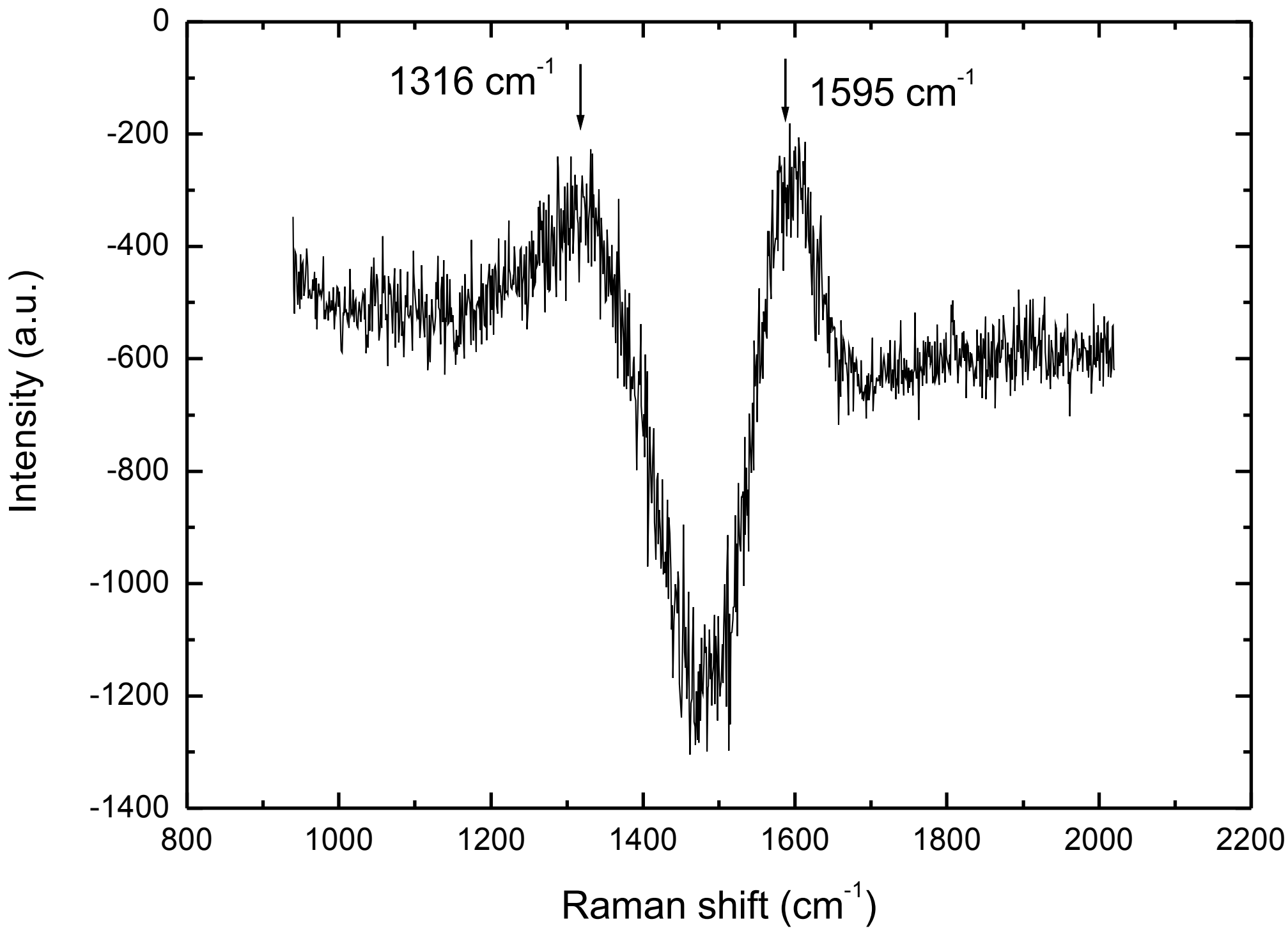

Fig. 3. The difference spectrum calculated from the spectra of Fig. 2.

The single and sharp second order Raman band (2D) at ~2670 $cm^{-1}$ has been widely used as a simple and efficient way to confirm the presence of single layer graphene. Raman spectrum of our sample in the 2D band regions is shown in Fig. 4. This spectrum is not sharp as expected and it can be fitted with two peaks at 2754 $cm^{-1}$ and 2853 $cm^{-1}$. This feature is typical for defected graphene. Such Raman spectra were already observed in the paper [13], where graphene was made luminescent. In this work graphene samples produced by microcleavage of graphite were exposed to oxygen/argon (1:2) RF plasma. The Raman spectra of the plasma treated graphene showed D, G, and broad second-order peak similar to ours. Similar short (1-6 s) was also the processing time.

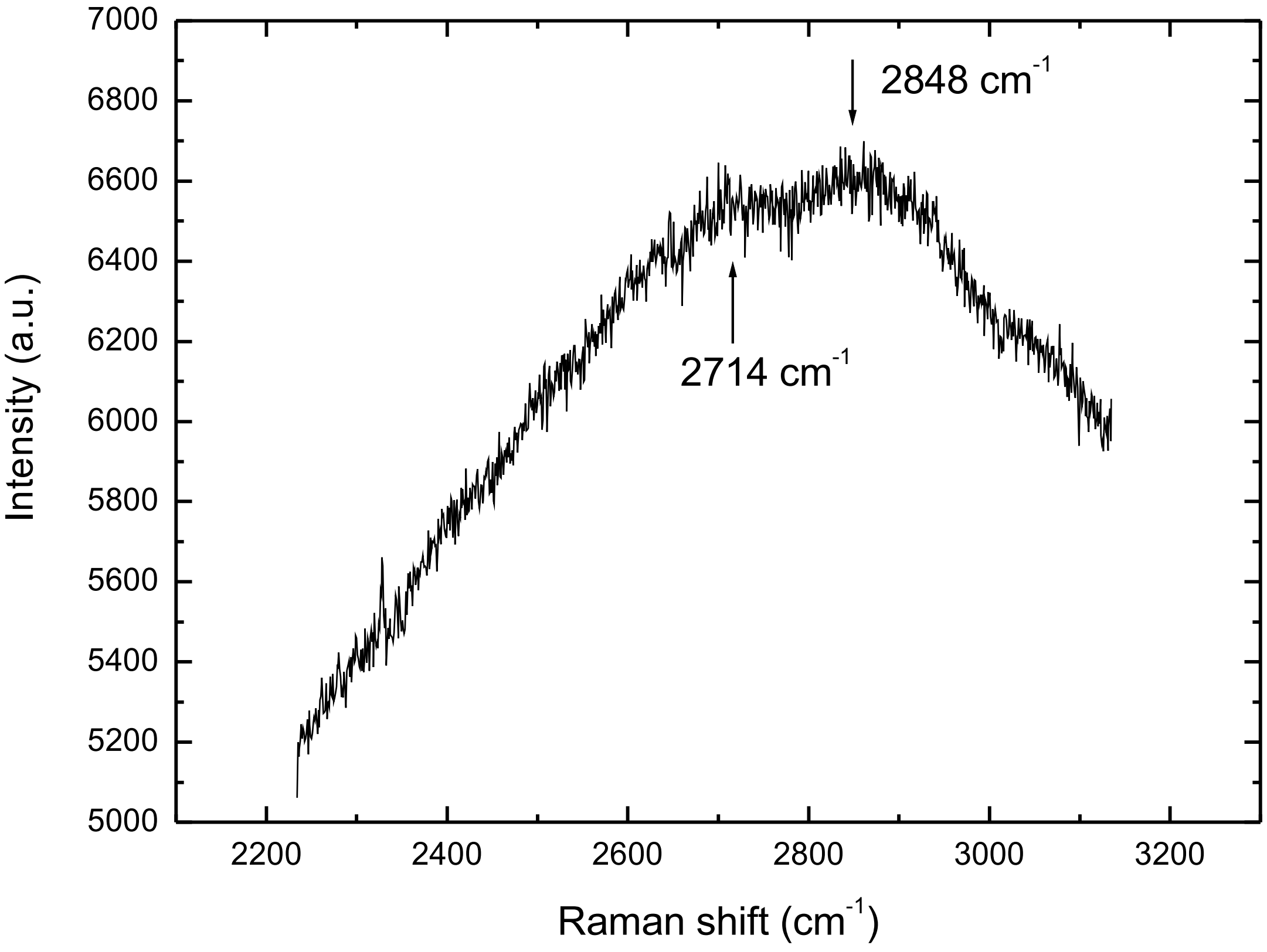

Fig. 4. Raman spectra of the modified film around the expected 2D band region of graphene.

Further the sample was annealed at 300ºC for 8 hours in order to enhance diffusion of the displaced carbon atoms in the modified surface layer. This temperature was chosen as maximal temperature which can not affect the underplaying DLC. Raman spectrum of the sample after annealing at 300 ºC for 8 h is shown on Fig. 5. The spectrum shows two distinct D- and G-peaks, which is typical for partial crystalline carbon with small crystalline size. Without ion bombardment such partial crystallization of DLC can be achieved by annealing at temperatures 800 - 900ºC. So the enhanced diffusion caused by the ion bombardment is evident. The second-order Raman peak, however, still remained broad.

Another sample (Fig. 6) was modified with 5 time higher dose of $2x10^{16}$ $Ar^{+}/cm^{2}$, but after the ion bombardment the sample was not annealed. Nevertheless similar Raman spectrum (Fig. 7) was observed - the same distinct D- and G-peaks and the broad 2D-peak on the right position.

One question arises immediately because the parameters used in this modification are typical for the material sputtering in such magnetron systems. Will the film not be milled by the argon ions during the modification? Indeed in spite of the very short processing time we have observed reduction of the film thickness. This can be seen in Fig. 7, where the strong Si-peak from the substrate at 520 $cm^{-1}$ is visible. In the virgin film (Fig. 6) this Si-peak was much smaller. Obviously the ion modification is a dynamical process, during which also film sputtering happens.

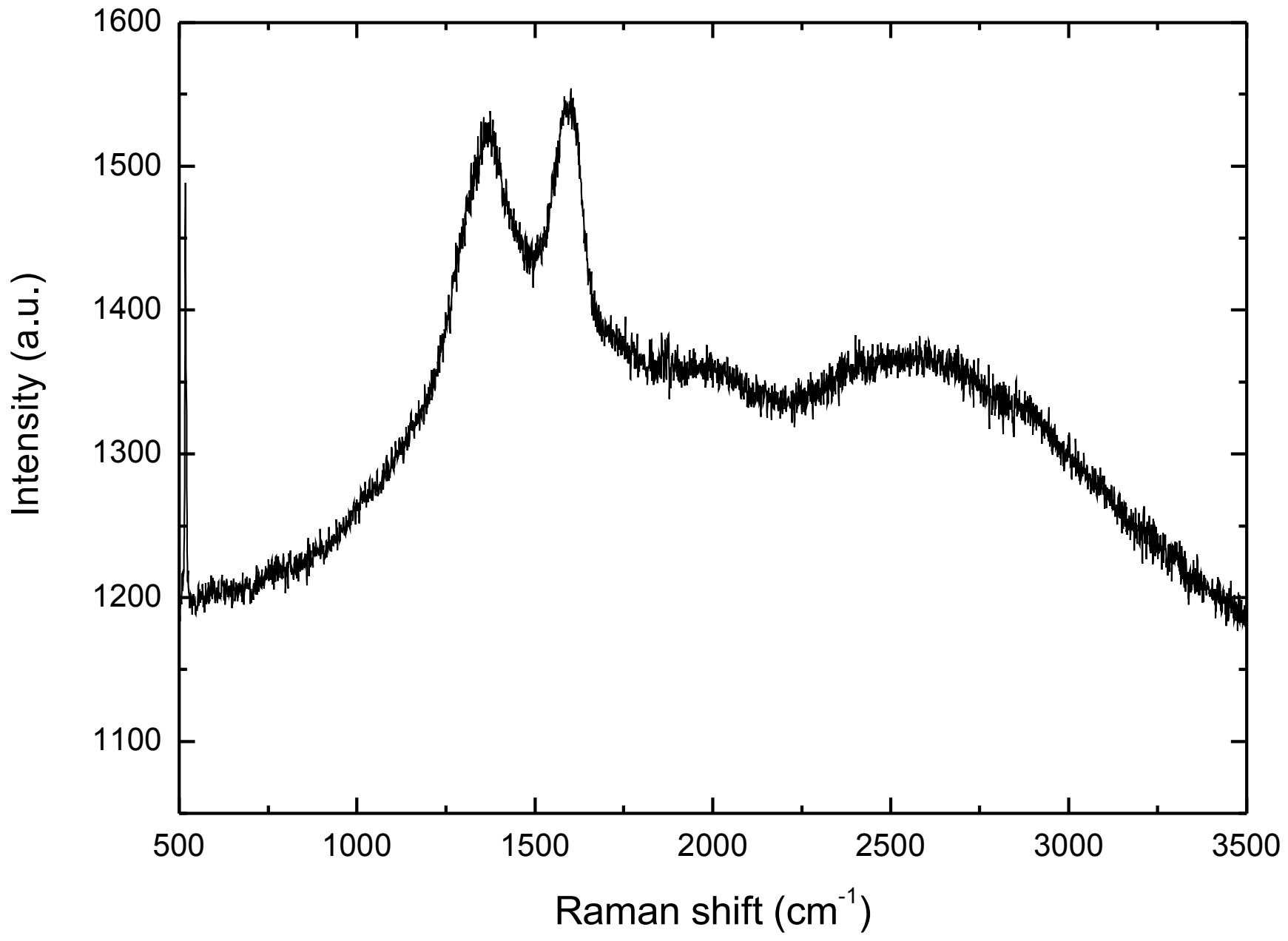

Fig. 5 Raman spectrum of the sample after annealing at 300 ºC for 8 h.

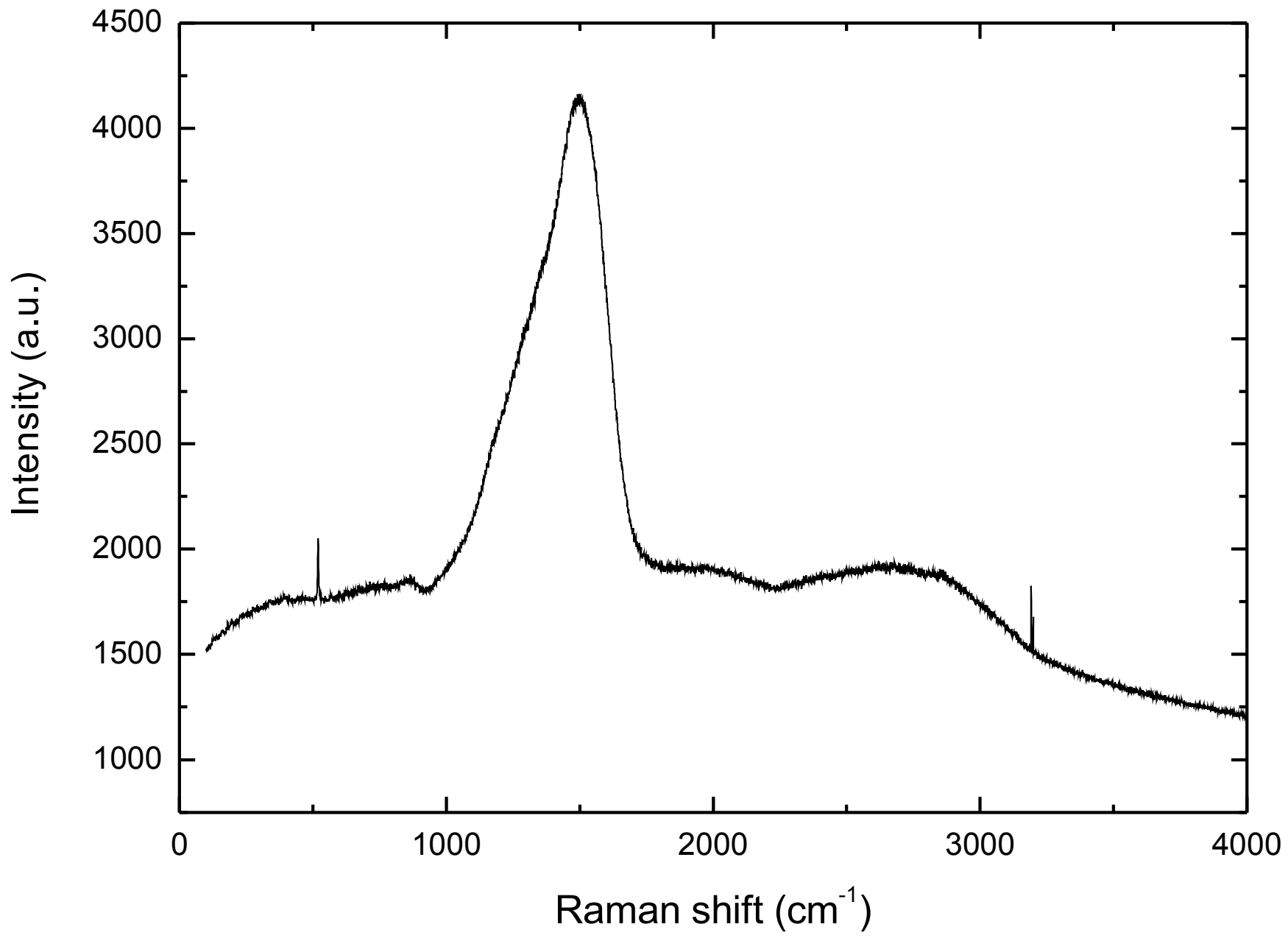

Fig. 6. Raman spectrum of the sample No. 2 before modification.

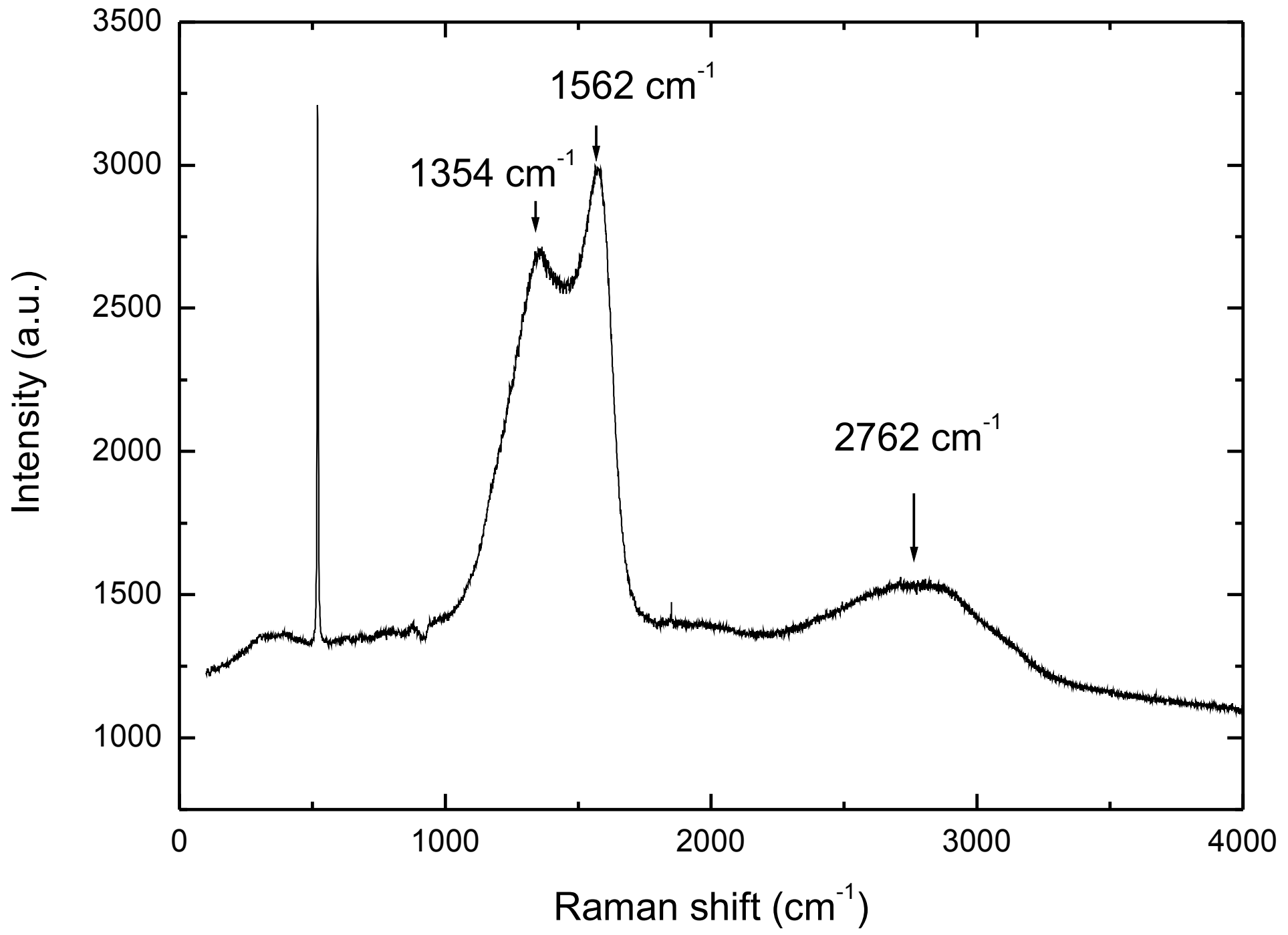

Fig. 7. Raman spectrum of sample No. 2 modified with dose at 2x$10^{16}$ $Ar^{+}/cm^{2}$ without annealing.

This result is very encouraging and we hope that by improving this technology it will be possible to fabricate defect-free graphene.

Acknowledgement: The authors would like to thank Dr. Evgenia Valcheva for the Raman measurements